\documentclass[prl,aps,twocolumn,footinbib,superscriptaddress,floatfix,bibliography]{revtex4-1}
\usepackage{graphicx}
\usepackage{amsmath}
\usepackage{comment}
\usepackage{bm}
\usepackage{hyperref}
\usepackage{color}

\usepackage{float}
\usepackage[caption=false]{subfig}	
\usepackage{tabularx,hhline}	
\usepackage{pstricks}
\usepackage{rotating}			       
\newcolumntype{P}[1]{>{\centering\arraybackslash}p{#1}}

\usepackage{bbding}
\usepackage{pifont}
\usepackage{wasysym}
\usepackage{amssymb}

\usepackage{textcomp}
\usepackage[T1]{fontenc}

\usepackage{tikz}
\usepackage{etoolbox}
\newrobustcmd*{\mysquare}[1]{\tikz{\filldraw[draw=#1,fill=#1] (0,0) rectangle (0.2cm,0.2cm);}}
\newrobustcmd*{\mycircle}[1]{\tikz{\filldraw[draw=#1,fill=#1] (0,0) circle [radius=0.1cm];}}
\newrobustcmd*{\mytriangle}[1]{\tikz{\filldraw[draw=#1,fill=#1] (0,0) -- (0.2cm,0) -- (0.1cm,0.2cm);}}
\newrobustcmd*{\myinvtriangle}[1]{\tikz{\filldraw[draw=#1,fill=#1] (0cm,0.2cm) -- (0.2cm,0.2cm) -- (0.1cm,0.0cm);}}
\newrobustcmd*{\mydiamond}[1]{\tikz{\filldraw[draw=#1,fill=#1] (0,0) -- (0.1cm,0.1cm) -- (0.0cm,0.2cm) -- (-0.1cm, 0.1cm); }}
\newrobustcmd*{\mypentagon}[1]{\tikz{\filldraw[draw=#1,fill=#1] (0,0) -- (0.05cm,0.0cm)--(0.1cm,0.1cm) -- (0.0cm,0.2cm) -- (-0.1cm, 0.1cm)--(-0.05cm,0.0cm)--(0.0cm,0.0cm);}}
\newrobustcmd*{\mysquaree}[1]{\tikz{\filldraw[draw=#1,fill=white, line width=0.7mm ] (0,0) rectangle (0.2cm,0.2cm) }}
\newrobustcmd*{\mycirclee}[1]{\tikz{\filldraw[draw=#1,fill=white, line width=0.7mm] (0,0) circle [radius=0.1cm];}}
\newrobustcmd*{\mytrianglee}[1]{\tikz{\filldraw[draw=#1,fill=white, line width=0.7mm] (0,0) -- (0.2cm,0) -- (0.1cm,0.2cm)--(0,0)--(0.2cm,0);}}
\newrobustcmd*{\myinvtrianglee}[1]{\tikz{\filldraw[draw=#1,fill=white, line width=0.7mm] (0cm,0.2cm) -- (0.2cm,0.2cm) -- (0.1cm,0.0cm)--(0cm,0.2cm);}}
\newrobustcmd*{\mydiamondd}[1]{\tikz{\filldraw[draw=#1,fill=white, line width=0.7mm] (0,0) -- (0.1cm,0.1cm) -- (0.0cm,0.2cm) -- (-0.1cm, 0.1cm)--(0,0)--(0.1cm,0.1cm); }}
\newrobustcmd*{\mypentagonn}[1]{\tikz{\filldraw[draw=#1,fill=white, line width=0.7mm] (0,0) -- (0.05cm,0.0cm)--(0.1cm,0.1cm) -- (0.0cm,0.2cm) -- (-0.1cm, 0.1cm)--(-0.05cm,0.0cm)--(0.0cm,0.0cm);}}
\newrobustcmd*{\mysquareee}[1]{\tikz{\filldraw[draw=#1,fill=white] (0,0) rectangle (0.2cm,0.2cm);}}
\newrobustcmd*{\mycircleee}[1]{\tikz{\filldraw[draw=#1,fill=white] (0,0) circle [radius=0.1cm];}}
\newrobustcmd*{\mytriangleee}[1]{\tikz{\filldraw[draw=#1,fill=white] (0,0) -- (0.2cm,0) -- (0.1cm,0.2cm)--(0cm,0cm);}}
\newrobustcmd*{\myinvtriangleee}[1]{\tikz{\filldraw[draw=#1,fill=white] (0cm,0.2cm) -- (0.2cm,0.2cm) -- (0.1cm,0.0cm) --(0cm,0.2cm);}}
\newrobustcmd*{\mydiamonddd}[1]{\tikz{\filldraw[draw=#1,fill=white] (0,0) -- (0.1cm,0.1cm) -- (0.0cm,0.2cm) -- (-0.1cm, 0.1cm)--(0,0); }}
\newrobustcmd*{\mypentagonnn}[1]{\tikz{\filldraw[draw=#1,fill=white] (0,0) -- (0.05cm,0.0cm)--(0.1cm,0.1cm) -- (0.0cm,0.2cm) -- (-0.1cm, 0.1cm)--(-0.05cm,0.0cm)--(0.0cm,0.0cm);}}
\definecolor{lightmagenta}{rgb}{0.9375,0.33203125,0.9375}
\definecolor{lightcoral}{rgb}{0.9375,0.5,0.5}
\definecolor{forestgreen(web)}{rgb}{0.13, 0.55, 0.13}
\definecolor{lightgreen}{rgb}{0.56, 0.93, 0.56}
\definecolor{royalblue(web)}{rgb}{0.25, 0.41, 0.88}
\definecolor{skyblue}{rgb}{0.53, 0.81, 0.92}
\definecolor{turquoise}{rgb}{0.19, 0.84, 0.78}
\definecolor{navyblue}{rgb}{0.0, 0.0, 0.5}
\definecolor{gray30}{rgb}{0.30078125,0.30078125,0.30078125}
\definecolor{gray80}{rgb}{0.796875,0.796875,0.796875}
\definecolor{slategray}{rgb}{0.44, 0.5, 0.56}
\definecolor{skyblue}{rgb}{0.53, 0.81, 0.92}
\definecolor{goldenrod}{rgb}{0.85, 0.65, 0.13}
\definecolor{brown(web)}{rgb}{0.65, 0.16, 0.16}

\graphicspath{{./figs/}} 		

\begin{document}

\title{Signatures of the finite-temperature mirror symmetry breaking in the $S$=1/2 Shastry-Sutherland model}
\author{Tokuro Shimokawa} 
\email{tokuro.shimokawa@oist.jp}
\affiliation{Okinawa Institute of Science and Technology Graduate University, 
Onna, Okinawa, 904-0495, Japan}


\date{\today}

\begin{abstract}

We investigate the finite-temperature properties of the $S$=1/2 Shastry-Sutherland Heisenberg model using a quantum typicality method. 
In the intermediate plaquette state region, we naturally expect to realize the finite-temperature phase transition associated with breaking the mirror symmetry of this model. We reveal some signatures of the spontaneous phase transition within a two-point correlation level at moderate temperatures since the constructed typical state can sense the existence of the degenerated excited states depending on the initial random state.
We also confirm that the local mirror order parameter shows the intriguing recovering phenomenon of the mirror symmetry in very low-temperatures, which could be understood from the nature of the ground and excited states of the finite-size systems. We expect that this recovering feature will disappear; instead, a saturated behavior appears in the local mirror order parameter in the thermodynamic limit. We discuss the relationship to the recent experimental results on ${\rm SrCu_2(BO_3)_2}$ under high pressures. 

\end{abstract}

\maketitle

\section{I. Introduction}
Quantum frustrated magnetism is well known as a source of the rich phenomena; however, it is still challenging to investigate its finite-temperature properties, significantly in more than one-dimensional systems.  One of the reasons for the difficulty is, the quantum Monte Carlo method, a powerful and nonbiased method, is generally not applicable to the quantum frustrated magnetism because of a negative sign problem. Although we can use another nonbiased method, the exact diagonalization (ED) method, this method is entirely limited to small-sized clusters. In this ED method, we need to use the canonical ensemble average, 
\begin{equation}
\langle \hat{A} \rangle^{\rm ens}_{\beta, N} = \sum_{\nu} \frac{ e^{-\beta E_{\nu}} }{Z(\beta)} \langle \nu | \hat{A} | \nu \rangle, 
\label{eq:ens}
\end{equation}
to get the expectation value of a physical quantity $\hat{A}$ at a temperature ${T=1/ \beta}$. Here, $E_{\nu}$ and $|\nu \rangle$ are the energy eigenvalues and the corresponding eigenvectors of the target Hamiltonian.The exponential increase of the memory cost for calculating eq.~({\ref{eq:ens}}) prevents us from treating large enough system size even if we use a modern supercomputer. For example, the number of spins we can treat is limited to about 24 in the case of an $S=1/2$ quantum spin system.  

For overcoming such the situation, alternative numerical methods without ensemble average ~\cite{Imada1986, Lloyd1988, Jakli1994, Hams2000, cTPQ, random2020} have also been developed and used historically to investigate the thermal nature of the quantum frustrated spin systems, such as triangular~\cite{Morita2020, Prelov2020, Seki2020, random2020}, kagome~\cite{cTPQ, Tomo2014, Shimokawa2016, Schnack2018, Hotta2018, Endo2018, Morita2020, Misawa2020, Patil2020, Prelov2020, Prelovsek2020, random2020}, and pyrochlore~\cite{Uematsu2019, Robin2020, Derzhko2020} magnets. In particular, the method using a typical pure state has got attention recently because it has been proven in different and independent literatures that a single pure state can represent the thermal equilibrium in the thermodynamic limit~\cite{Lloyd1988, Hams2000, cTPQ, random2020}.
This typicality method can dramatically reduce the computational cost; more concretely, we can handle the twice large system compared to the ED method because the memory cost is reduced from $O(2^N \times 2^N)$ to $O(2^N)$ when our target system is constructed by interacting $N$ $S=1/2$ spins.

However, whether this method can capture finite-temperature phase transitions is still a challenging problem because of the huge finite size effect near the critical point, although it can handle twice as large as sizes. We may not be able to see a signature of the finite-temperature phase transition, for example, as the peak of the specific heat. It is interesting and useful to know whether we can detect some other types of the sign of the phase transition from the typicality method.

In this paper, we will address this question with a concrete example, the $S=1/2$ Shastry-Sutherland model~\cite{SS1981}, and will show a practical way via the combined the typicality and the exact diagonalization methods to see a clear signature of the finite-temperature phase transition between paramagnetic state and mirror symmetry broken state within a two-point correlation level.

\begin{figure}[b]
  \includegraphics[clip,width=7.0cm]{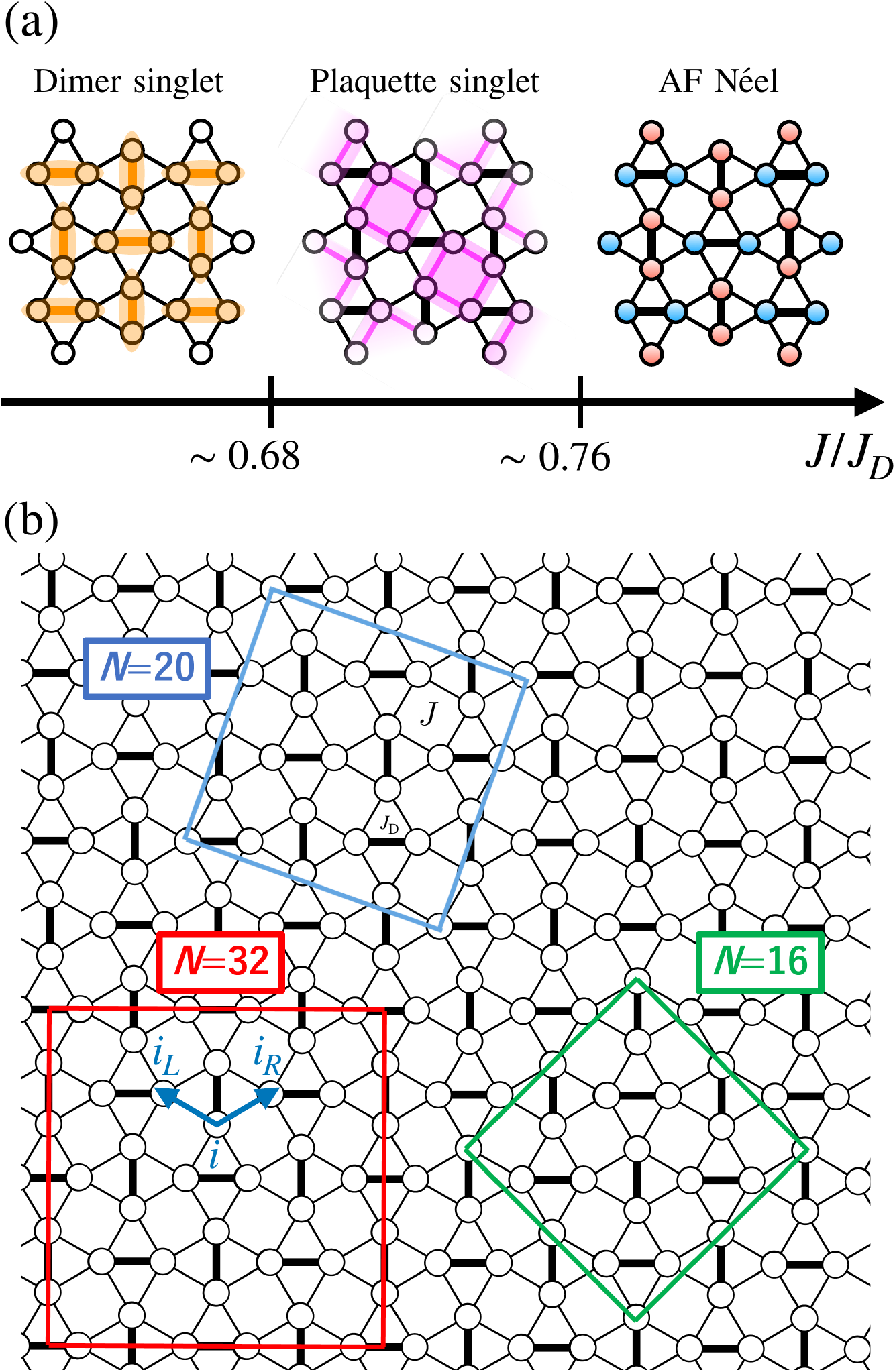}
      \caption{(a)The ground state phase diagram. (b) Finite-size clusters we treat in this study. $J$ and $J_D$ are inter and intra dimer interactions. The site labels, $i$, $i_L$, and $i_R$ are used in the local mirror order parameter of eq.~(\ref{eq:local}) to detect the empty plaquette state.}
    \label{fig:GS-cluster}
    \end{figure}

We will investigate the low-temperature properties of the $S=1/2$ Shastry-Sutherland Heisenberg antiferromagnet~\cite{SS1981} via the typicality method, especially focusing on the intermediate plaquette singlet region. This famous model is known as a two-dimensional orthogonal dimer model and was originally proposed by Shastry and Sutherland in 1981~\cite{SS1981}. The Hamiltonian is defined as follows,

\begin{equation}
\mathcal{H}=J_D \sum_{\langle i,j \rangle} {\bf S}_i \cdot {\bf S}_j + J \sum_{\langle \langle i,j \rangle \rangle} {\bf S}_i \cdot {\bf S}_j .
\label{eq:Ham}
\end{equation}
The schematic view of the model and the ground state phase diagram are shown in Fig.~\ref{fig:GS-cluster}.  Here, $J_D$ and $J$ are intra- and inter-dimer exchange couplings. The ground state of this model has been investigated well~\cite{SS1981, Miyahara1999, Koga2000, Zheng2001, Chung2001, Lauchli2002, Miyahara2003, Lou2012,Corboz2013, Nakano2018}; the dimer-singlet state for $J/J_D<0.68$ and the N${\rm \acute{e}}$el state for $J/J_D>0.76$, the intermediate plaquette singlet state appears 0.68<$J/J_D$<0.76. 

    Fortunately this model has a realization material,  ${\rm SrCu_2(BO_3)_2}$~\cite{Kageyama1999, Kageyama2000}. It was reported that this material lies in the dimer product state, and the recent remarkable progress in experiments is that high pressure makes this material into the plaquette singlet state with increasing the interaction ratio $J/J_D$ \cite{Kageyama2003, Waki2007, Takigawa2010, Haravifard2012, Ohta2015, Haravifard2016, Zayed2017, Sakurai2018, Bettler2020, Guo2020, Jimenez2020}. It is well known that the Mermin-Wagner theorem does not prohibit the finite-temperature phase transition associated with breaking the discrete mirror symmetry in the plaquette singlet region. Indeed, two independent groups reported quite recently the presence of a small peak of the specific heat at a low temperature, $\sim$ 2K, under high pressure around 22 kbar, which may be considered as the finite-temperature phase transition from the paramagnetic state to the intermediate plaquette singlet state \cite{Guo2020, Jimenez2020}.  They also conducted the full exact diagonalization calculation based on small size clusters up to 20 spins and a sophisticated iPEPS calculation to investigate the thermal properties of this material; however, detecting the phase transition at very low-temperatures seems to be hard for both methods. Another theoretical study is strongly required.

This article is organized as follows: Sec. II is for the details of our numerical methods. In Sec. III, we show the temperature dependencies of the specific heat, the real-space two-point correlation functions, and the local mirror order parameter computed by the typicality method. We also show the low energy spectrum obtained by the thick-restarted Lanczos method for understanding the finite-temperature property. Sec. IV and V are devoted to discussion and summary.

\section{II. The numerical methods}

We use the Sugiura and Shimizu method to construct a set of typical pure states for investigating the thermal properties of the $S=1/2$ Shastry-Sutherland model, eq.~({\ref{eq:Ham}}). We briefly describe the detailed procedure in this method. According to the ref.~\cite{cTPQ} , a set of typical pure states,  $|\beta, N \rangle$, for the inversed temperature $\beta=1/T$ and the system size $N$ can be constructed operating our target Hamiltonian on a set of initial random vectors,

\begin{equation}
|\beta, N\rangle=e^{-\beta \mathcal{H}/2}|\psi_0\rangle,
\label{eq:PureState}
\end{equation}
where the initial random vectors are given by $|\psi_0\rangle =\sum_{i=1}^{2^N} c_i | i \rangle$. 
Here, ${\{c_i\}}$ are random complex numbers satisfying the normalization condition $\sum_{i=1}^{2^N} |c_i|^2=1$ and $\{| i \rangle \}$ are an arbitrary orthonormal basis set of the Hilbert space of $\mathcal{H}$. 
We use a binary bits representation for the orthonormal basis set such as $| i \rangle =|\sigma_1 \sigma_2 ... \sigma_N \rangle$ with $\sigma_j=\uparrow$ or $\downarrow$.  Note that we use the U(1) symmetry of the Hamiltonian for reducing the memory cost of our computation, and the eq.~(\ref{eq:PureState}) can be divided by the quantum conserved number, $S_z^{\rm tot}$.

Sugiura and Shimizu call a $|\beta, N\rangle$ as a canonical thermal pure quantum (TPQ) state which represents an equilibrium state.
The ensemble average of a physical quantity $\hat{A}$, eq.~(\ref{eq:ens}) could be replaced by the following equation,

\begin{equation}
\langle \hat{A} \rangle^{\rm TPQ}_{\beta, N}= \frac{ \overline{ \langle \beta, N| \hat{A} |\beta, N \rangle}} { \overline{\langle \beta, N  |\beta, N\rangle}},
\label{eq:TPQave}
\end{equation}
where the overline denotes the average over the initial random vectors.
We should note that we can get the correct value from one TPQ state, that means, $\langle \hat{A} \rangle^{\rm TPQ}_{\beta, N}=\langle \hat{A} \rangle^{\rm ens}_{\beta, N}$,  only when we treat large enough system size.
For small system sizes, we need to average many realizations of the random initial vectors.
It was shown in many pioneering and independent researches \cite{Imada1986, Lloyd1988, Jakli1994, Hams2000, cTPQ} that the deviation $\overline{(\langle \hat{A} \rangle^{\rm TPQ}_{\beta, N} - \langle \hat{A} \rangle^{\rm ens}_{\beta, N} )^2}$ decays as $\sim 1/\sqrt{ID}$ where the $I$ means the number of realizations of the initial random vectors, and the $D$ represents the dimension of the entire Hilbert space of the model. For reducing the deviation from the exact value, in our computation, the average over initial random vectors is taken over 800 ($N$=16), 100 ($N$=20), and 20 ($N$=32) realizations, and we evaluate the error of each physical quantity by the standard error.

We also use the thick-restart Lanczos algorithm \cite{Thick2000, QS3} for obtaining the ground and excited states of our target Hamiltonian. Despite the conventional Lanczos algorithm, this variant Lanczos algorithm enables us to avoid losing orthogonality, which results in spurious eigenvalues and eigenvectors.

\section{III. Results}
\begin{figure}[htbp]
  \includegraphics[clip,width=6.0cm]{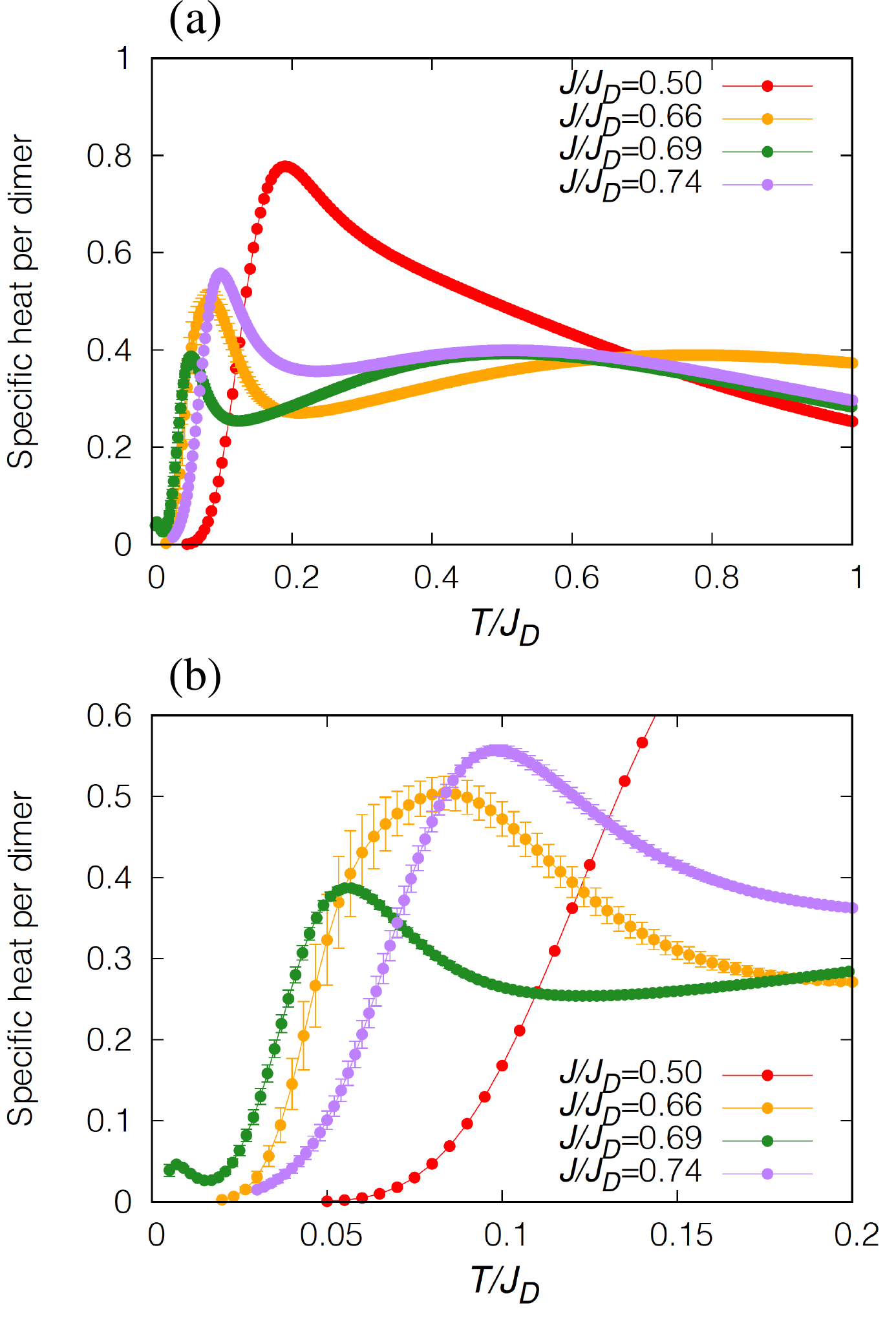}
      \caption{The $N$=32 temperature dependence of the specific heat per dimer in $J/J_D$=0.5, 0.66, 0.69 and 0.74 for (a) $T/J_D\leq 1$ and (b) $T/J_D\leq0.2$. }
    \label{fig:heat}
    \end{figure}

The TPQ calculation results of the specific heat in several $J/J_D$ values are shown in Fig.~\ref{fig:heat} in the temperature regions, (a) $T/J_D \le 1.0$ and (b) $T/J_D \le 0.2$, respectively. We here focus on our maximum finite-size cluster, $N$=32. We confirm that our computations in $J/J_D$=0.50 and 0.66 are consistent with the previous QMC, and TPQ works in ref.~\cite{Wessel2018, WietekTPQ2019}.
Let us focus on the data in $J/J_D$=0.69 here, exhibiting two peak structures at $T/J_D \sim 0.05$ and $T/J_D \sim 0.007$. 
The finite-temperature iPEPS calculation in this parameter was done quite recently~\cite{Jimenez2020} and exhibited the peak structure only at $T/J_D \sim 0.05$. We also find an additional small peak at very low temperatures, $T/J_D \sim 0.007$ in the $N=32$ cluster. 
We note that this low-temperature small peak appears only in the intermediate plaquette singlet phase and near the phase boundary between the dimer singlet and plaquette singlet phases, $J/J_D<0.74$, therefore, this small peak corresponds to the Schottky anomaly which was not confirmed in small size ED calculations in ref.~\cite{Guo2020}.

\begin{figure*}[htbp]
  \includegraphics[clip,width=15.5cm]{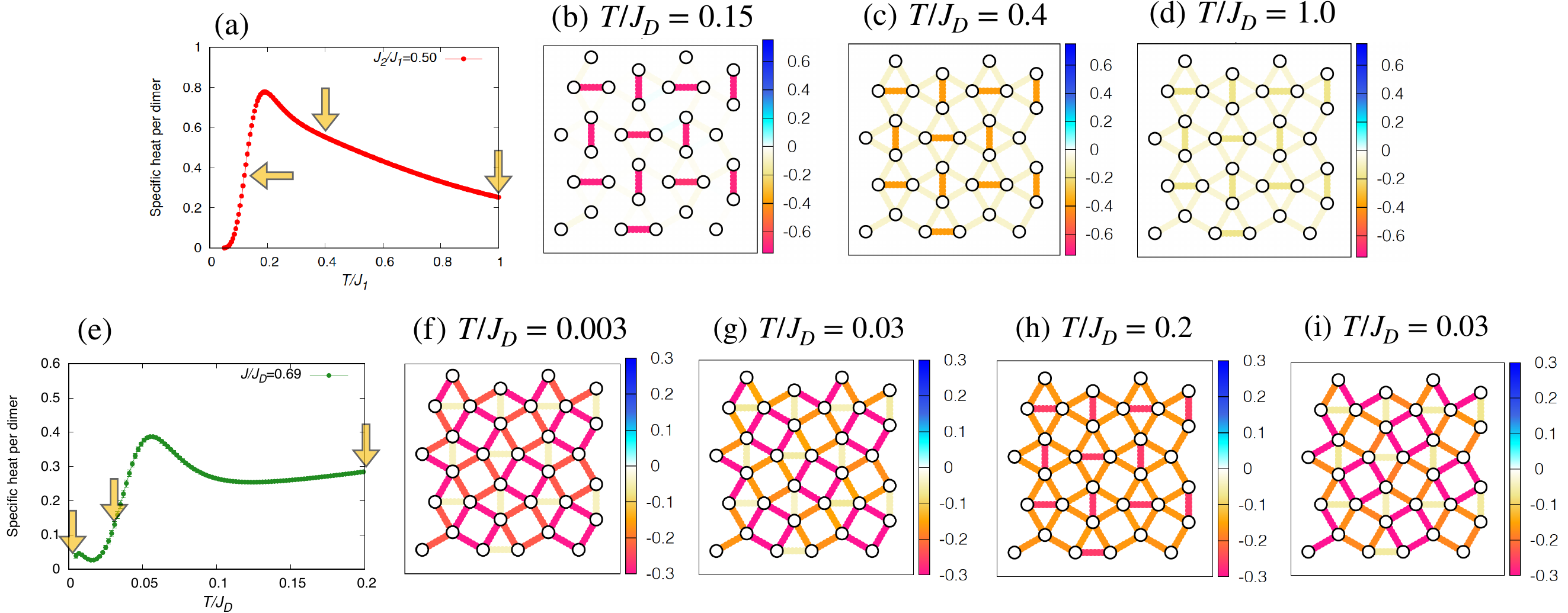}
      \caption{The temperature dependencies of the specific heat per dimer and the nearest-neighbor two-point correlation function in the $N=32$ cluster. The upper (lower) panels show the results in $J/J_D$=0.50 (0.69) as the typical example in the dimer (empty plaquette) regions. The orange arrows point to the temperatures we treated for the real-space correlation functions. The color in each nearest-neighbor bond in (b-d) and (f-i) represents the intensity plot of the corresponding two-point correlation function ${\bf S}_i \cdot {\bf S}_j$. The results in (b-d) and also in (f-h) are obtained through a random initial vector, respectively. The result in (i) is obtained from a different initial vector from that used in (f-h).}
    \label{fig:RealCorr}
    \end{figure*}

To investigate thermal development in correlations, we calculate the temperature dependence of the real-space nearest-neighbor two-point correlation function. We show the $J/J_D=0.5$ results for the dimer-singlet phase in Fig.~\ref{fig:RealCorr}~(b-d), and the $J/J_D=0.69$ results for the singlet-plaquette phase near the phase boundary in Fig.~\ref{fig:RealCorr}~(f-h). Note that we show the real-space correlations obtained from an initial random vector without taking the average over it. In the dimer-singlet phase, we just could see the enhancement of the intra-dimer AF correlation with decreasing the temperature from Fig.~\ref{fig:RealCorr}~(d) to (b), and this behavior does not change qualitatively in the results obtained from different initial random vectors. In the intermediate plaquette phase, on the other hand, we can see a clear signature of a mirror symmetry breaking pattern at moderate temperatures, $0.007 \lesssim T/J_D \lesssim 0.05$, as shown in Fig.~\ref{fig:RealCorr}~(g). There are two choices for the empty plaquette pattern, and we confirm which pattern appears depends on the initial random vectors (see also Fig.~\ref{fig:RealCorr}~(i)). One could also see a more intriguing feature, the recovering of the mirror symmetry, at lower temperatures, $T/J_D \lesssim 0.007$ as shown in Fig.~\ref{fig:RealCorr}~(f). 
We here also comment that the multiple peak structures in the specific heat in each parameter are associated with the change in the observed correlation patterns.

\begin{figure}[htbp]
  \includegraphics[clip,width=5.7cm]{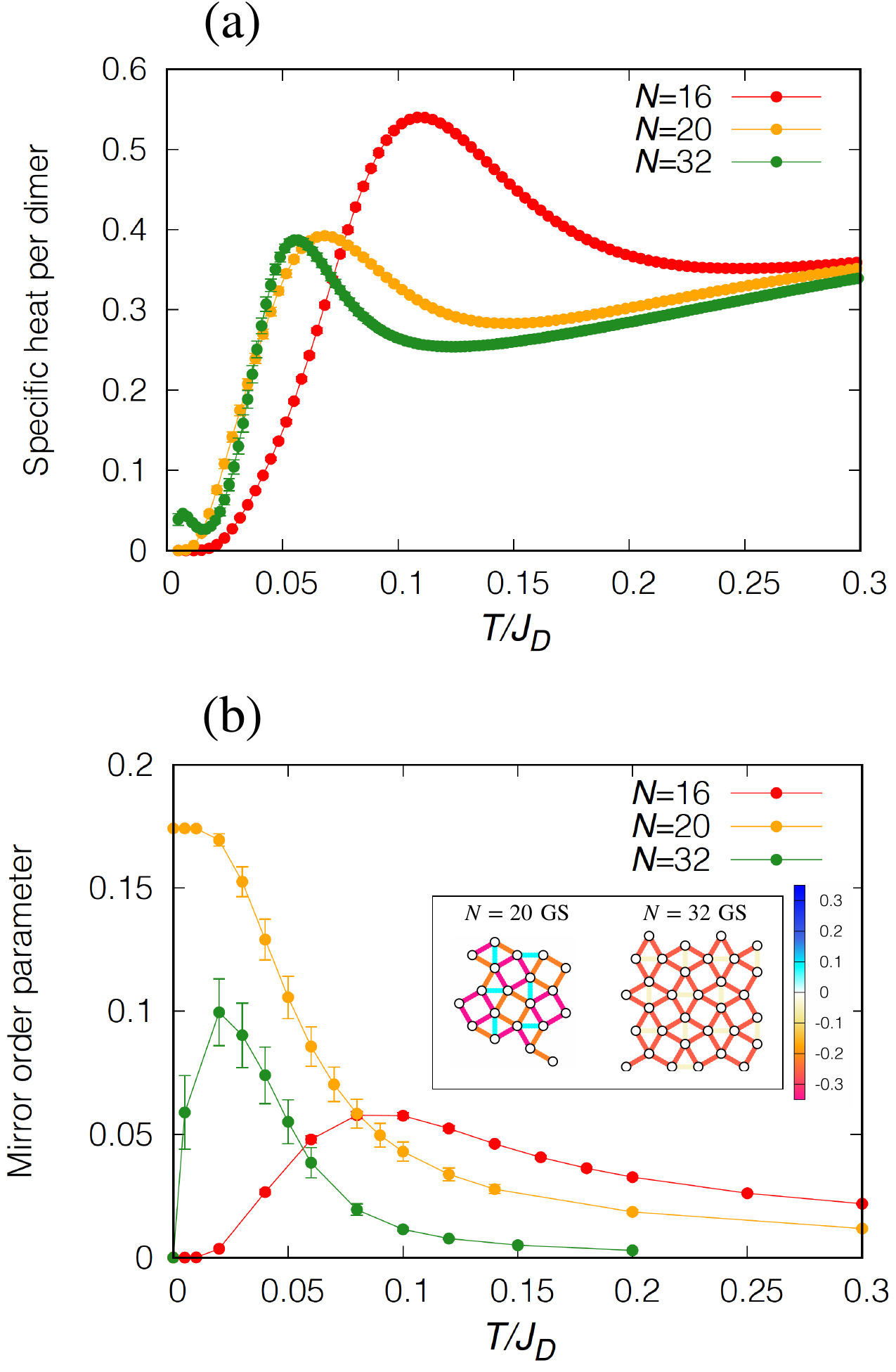}
      \caption{The size and temperature dependencies of (a) the specific heat and (b) the local mirror order parameter in $J/J_D$=0.69. The inset of (b) show the results of the real-space nearest-neighbor correlation function in the ground state of $N=20$ (left) and $32$ (right) in $J/J_D$=0.69.}
    \label{fig:heat-local}
    \end{figure}

We investigate more details focusing on the $J/J_D$=0.69 case.
We will see the size dependencies of the specific heat and of the following local mirror order parameter,

\begin{equation}
\frac{1}{N_m}\sum_{i=1}^{N_m} \langle |{{\bf S}_i \cdot {\bf S}_{i_L}- {\bf S}_i \cdot {\bf S}_{i_R}} |\rangle^{\rm TPQ}_{\beta, N},
\label{eq:local}
\end{equation}
where the $i$ means the lower site on the vertical $J_D$-bond, $N_m$ is the number of site $i$, which is taken as $N_m=N/4$, and 
$i_L$ and $i_R$ are the left and right-hand side sites connected to the $i$-th site (For example, see also the $N$=32 cluster in Fig.~\ref{fig:GS-cluster}~(b)). 
In Fig.~\ref{fig:heat-local}, finite-size effects appear in the specific heat and the local mirror parameter at a lower temperature than $T/J_D$=0.3. 
However, the higher temperature peak in the specific heat tends to shift to a lower temperature with the size and almost seems to be converged. 
Besides, we confirm the consistency with the iPEPS calculation in this peak temperature; therefore, we may expect that the $N$=32 cluster is not so far from the thermodynamic limit, at least at $T/J_D\gtrsim0.05$. 

We could find a bit complicated finite-size dependence of the local mirror operator in Fig.~\ref{fig:heat-local}~(b). 
Although the mirror symmetry recovering features are obtained in the $N$=16 and 32 clusters as we confirmed in Fig.~\ref{fig:RealCorr}~(f-h), the 20-site cluster does not show it, and the value of the eq.~(\ref{eq:local}) saturates to a nonzero constant value in $T \rightarrow 0$.
This complicated finite-size dependence could be understood from the ground state property of each cluster.
For example, in the $N$=32 cluster, the ref.~\cite{Lauchli2002} reported that the ground state is constructed by an s-wave type wave function exhibiting a plaquette type dimer-dimer correlation pattern. Note that this type of correlation pattern can not be observed within the two-point correlation function level because the current cluster has mirror symmetry and the ground state is not degenerated. Indeed, we confirmed this fact by calculating the nearest neighbor two-point correlation in the ground state of the $N=32$, which is shown in the inset of Fig.~\ref{fig:heat-local}~(b). We can see clearly that the eq. (5) should be zero in this ground state. The same situation also happens in the $N=16$ case.
In contrast, the $N$=20 cluster also has a unique ground state but does not have the mirror symmetry in itself, which means there is only one empty plaquette pattern in this cluster, which is detectable in two-point correlation function level as shown in the inset of Fig.~\ref{fig:heat-local}~(b).

\begin{figure}[t]
  \includegraphics[clip,width=8.5cm]{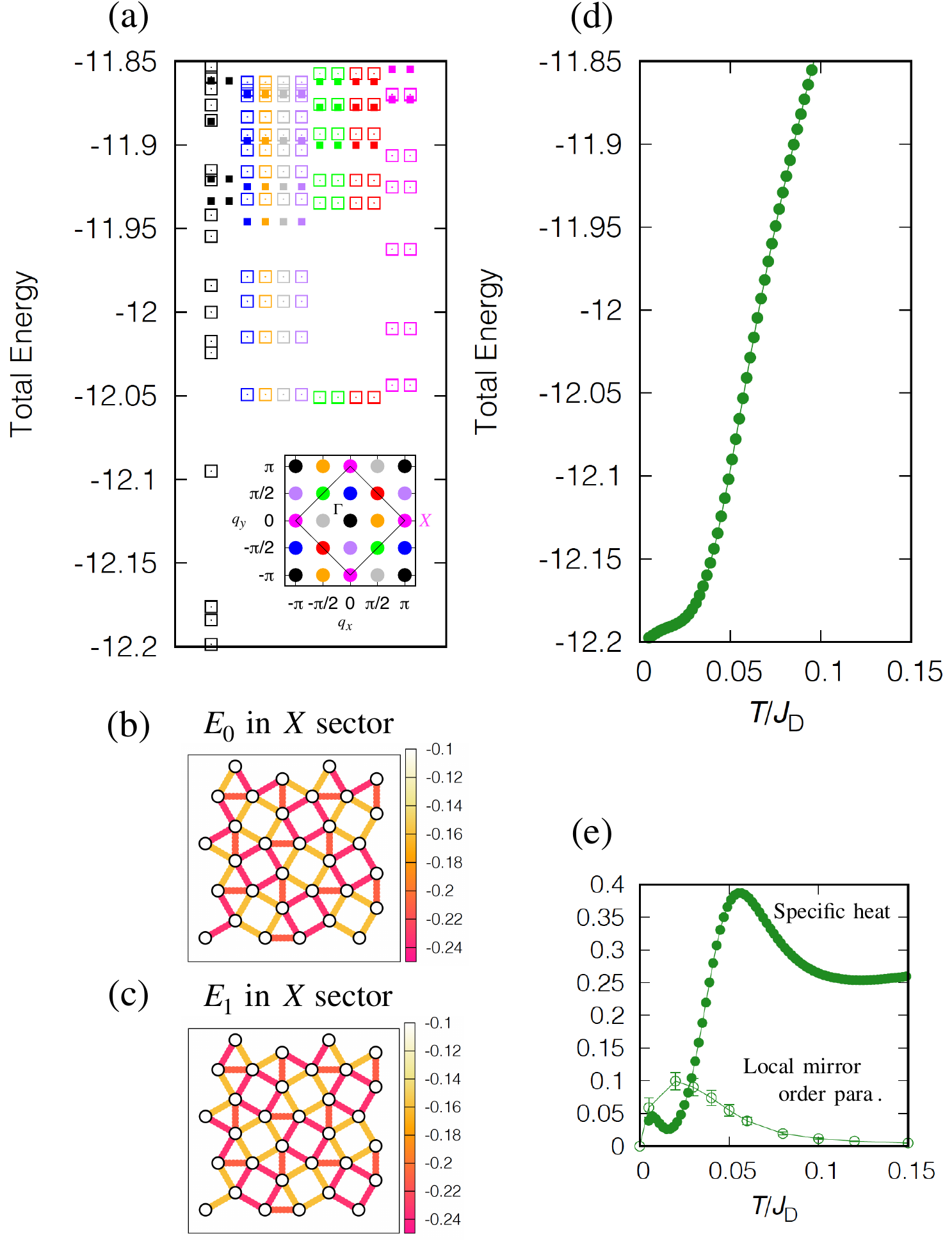}
      \caption{(a) Energy spectrum in $N$=32 and $J/J_D$=0.69 obtained by the thick-restarted Lanczos method. The empty (filled) symbols show the results of total S=0 (1). The inset shows the momentum sector. (b-c) The real-space nearest-neighbor correlation function obtained in the lowest two states in the $X$ sector. (d-e) The temperature dependencies of the total energy, the specific heat per dimer, and the local mirror order parameter in $N=32$ and $J/J_D=0.69$ obtained by our typicality method.}
    \label{fig:Spectrum}
    \end{figure}

One may have additional questions at the moment; why could we see the symmetry broken feature within the two-point correlation level at moderate temperatures in $N$=16 and 32 clusters? What happened in the TPQ state constructed by the eq.~(\ref{eq:PureState})?
To answer these questions, we calculate the low-energy states by mean of the thick-restarted Lanczos method using U(1) and translational invariances and reveal the existence of the two-fold degenerated excited states exhibiting mirror symmetry breaking pattern in two-point correlation function, which belong to the momentum sectors on the boundary line of the 1st Brillouin Zone (see the red, green, magenta points in Fig.~\ref{fig:Spectrum}~(a)).
For example, we show that the lowest two-fold states on $X$ point can exhibit clear empty plaquette type correlations as shown in Fig.~\ref{fig:Spectrum} (b-c).

Let us discuss here the origin of the mirror symmetry breaking feature at moderate temperatures. According to the relationship between the eq.~(\ref{eq:ens}) and eq.~(\ref{eq:TPQave}), we can expect that a normalized constructed typical state by an initial random vector is written by the eigenvalues and eigenvectors of the Hamiltonian~\cite{Yamaji2018},

\begin{equation}
\frac{|\beta, N\rangle}{\sqrt{\langle \beta, N|\beta, N\rangle}}\sim \sum_{\nu} \exp{(i \phi_\nu)}  \frac{e^{-\beta/2 E_\nu}}{\sqrt{Z(\beta)}} |\nu \rangle,
\label{eq:PureState2}
\end{equation}
where the $\phi_\nu$ corresponds the phase degrees of freedom.
In our typicality method, it is expected that this $\phi_\nu$ and also how the degenerated states are constructed are completely determined by initial random complex numbers $ \{c_i\} $ which are used for the initial random vector, $|\Phi_0\rangle$ of eq.~(\ref{eq:PureState}).
This is the reason why we could see the two patterns of the empty plaquette in the real-space two-point correlation function with depending on the initial random vectors as shown in Fig.~\ref{fig:RealCorr} (g) and (i). We also calculate the temperature dependence of the total energy, and the result shows that when our system reaches the total energy level lower than $\sim-11.85$ at the lower temperature than $T/J_D \sim 0.1$, the local mirror order parameter starts to be enhanced as shown in Fig.~\ref{fig:Spectrum}~(a) and (d-e). These results indicate that the typical state can sense the existence of the degenerated states, and show the mirror symmetry breaking pattern at moderate temperatures with fully depending on the initial random vectors. If we could treat much larger systems by means of the typicality method, we expect that one quantum typical state will select an empty plaquette pattern at moderate temperature, which can be considered as a {\it spontaneous} finite-temperature phase transition.

\section{IV. Discussion}
Let's discuss the fate of the local mirror order parameter and the specific heat in the thermodynamic limit in the intermediate plaquette region.
According to the paper by L${\rm \ddot a}$uchli et al~\cite{Lauchli2002}, the ground state and the 1st excited state exhibiting the empty plaquette dimer-dimer correlation function is expected to be degenerated in the thermodynamic limit, and then can show the plaquette pattern even in the two-point correlation level.
Therefore, the low-temperature recovering feature of the mirror symmetry is just a finite-size effect and we can expect that a saturated behavior will appear in the local mirror order parameter at $T \rightarrow 0$ in much larger system sizes.
Our computational results in $N$=16 and 32 in Fig.~\ref{fig:heat-local}~(b) may support our scenario because the peak temperatures of the local mirror order parameter is shifted to a lower temperature and seems to exhibit a saturated behavior in the larger system size. In much larger system sizes, we may expect the existence of the additional sharp peak of the specific heat which indicates the Ising-type second order phase transition, however, this is very difficult to detect in our typicality computation~\cite{Guo2020,memo}.

We comment about the relationship to the recent experimental results on ${\rm SrCu_2(BO_3)_2}$. The NMR and the recent inelastic neutron scattering (INS) measurements~\cite{Waki2007, Zayed2017} reported that this material under high pressure did not give the empty plaquette state, but the full plaquette state in which the singlet is formed in each $J$ square plaquette with $J_D$ interaction. The discrepancy from the simple Shastry-Sutherland model has not been settled and the distortion~\cite{Boos2019}, Dzyaloshinkii-Moriya~\cite{DMint}, and 3D coupling~\cite{Wessel2018} interactions were proposed as the origin of it.
The important thing is that both empty and full plaquette states are two-fold degenerated ground states (if there is no distortion); therefore, we can expect that the almost same physics appears at finite temperatures as we observed in our study. In connection with this, we also compute the temperature dependence of another local mirror order parameter which can detect the mirror symmetry breaking feature in the full plaquette manner, and confirm that this full plaquette correlation is also enhanced at moderate temperatures although the dominant correlation is empty one (see also our Appendix). Whether additional interactions to our present Hamiltonian can change the dominant correlation at moderate temperatures is an interesting problem for ${\rm SrCu_2(BO_3)_2}$. As we noted above, it is expected that the finite-size effect is not so strong around $T/J_D \sim 0.05$ even in the $N$=32 cluster. Computing dynamical physical quantities using recent new techniques based on the typicality method~\cite{Ikeuchi2015, Hotta2018, Yamaji2018, random2020} around the temperature, having finite values in the local mirror order parameter, also can give useful information for the NMR, INS, and ESR measurements.

Finally, we hope that the relationship between the signature of the phase transition and the existence of the corresponding degenerated excited states can open a new door via the typicality method for understanding the nature of more complex finite-temperature phase transitions such as a topological phase transition associated with topological defects.

\section{V. Summary}
By the quantum typicality method, we could confirm the clear signature of the mirror symmetry breaking pattern at low temperatures in the $S$=1/2 Shastry-Sutherland model. The signature in the two-point correlation level comes from the existence of the degenerated excited states having empty plaquette correlations, and the typical state can sense them depending on the random initial states. The recovering feature of the mirror symmetry is a finite-size effect, and we could expect a saturated behavior in the local mirror order parameter in the thermodynamic limit.
\vspace{\baselineskip}

\subsection{\label{sec:ACKNOWLEDGMENTS}ACKNOWLEDGMENTS}

The author thanks Shunsuke Furuya, Eiki Iyoda, Tsutomu Momoi, Tsuyoshi Okubo, Daisuke Yamamoto, Hiroshi Ueda, Karlo Penc, Matthias Gohlke and Nic Shannon for fruitful discussions. 
This work is supported by the Theory of Quantum Matter Unit of the 
Okinawa Institute of Science and Technology Graduate University (OIST) and also supported by JSPS KAKENHI Grant Number 19K14665.
Our TPQ code is based on TITPACK ver. 2 and the reliability was checked by HPhi application~\cite{Kawamura2017}. 
The thick-restarted Lanczos code is based on ${\rm QS^3}$~\cite{QS3}.
Parts of the numerical calculations are performed using the facilities of the 
Supercomputing Center, ISSP, the University of Tokyo, and OIST.

\appendix
\section{APPENDIX}

\begin{figure}[b]
  \includegraphics[clip,width=8.5cm]{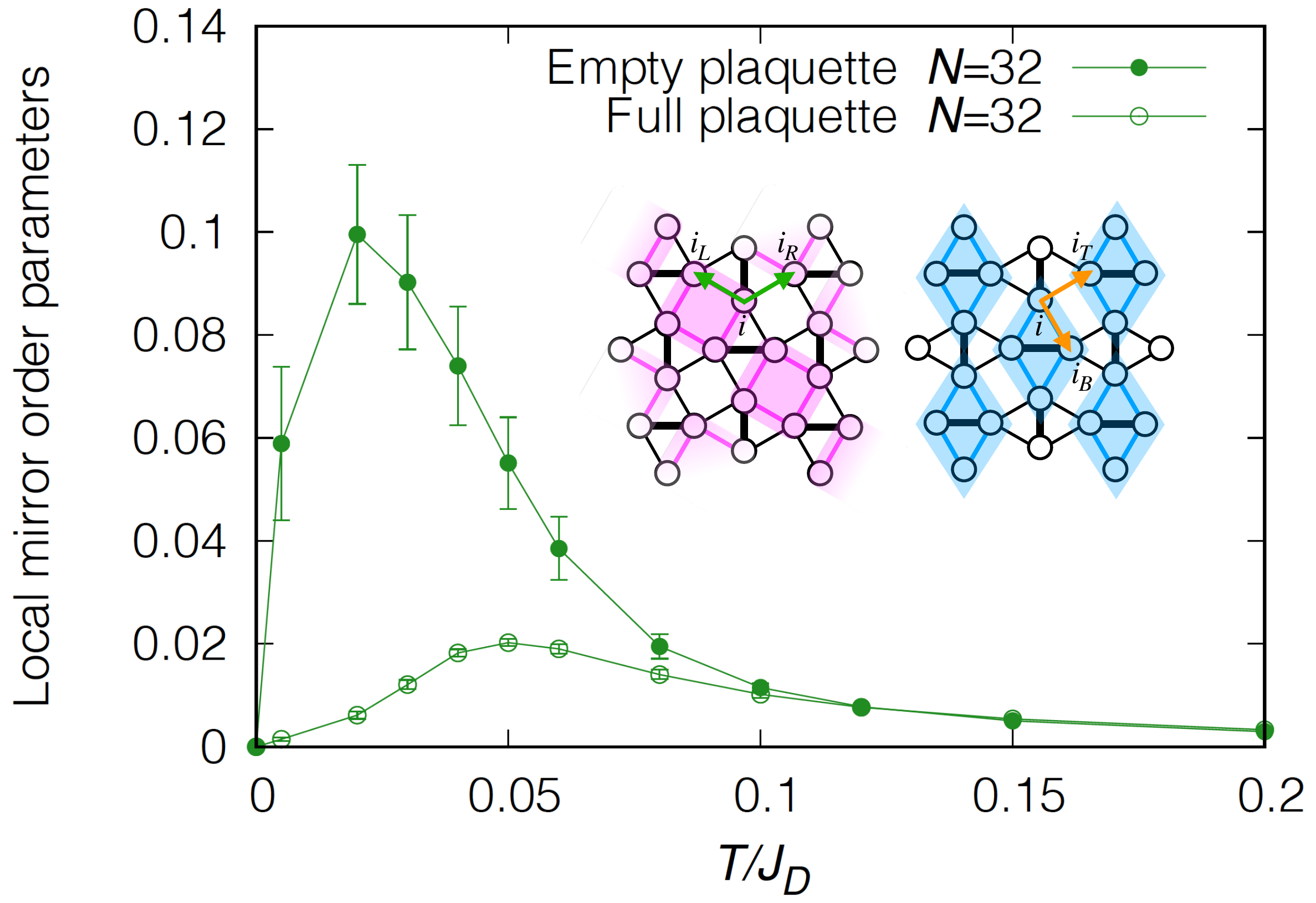}
      \caption{The temperature dependencies of the local mirror order parameters for empty plaquette and full plaquette in $N=32$ and $J/J_D$=0.69. The inset graphs show the empty (left) and full (right) plaquette pictures and the the labels in these pictures show the site information used for the corresponding local mirror order parameters.}
    \label{fig:full}
    \end{figure}

In this Appendix, we treat another type of local mirror order parameter for the full plaquette state.
The local mirror order parameter for the full plaquette state is defined as
\begin{equation}
\frac{1}{N_m}\sum_{i=1}^{N_m} \langle |{{\bf S}_i \cdot {\bf S}_{i_T}- {\bf S}_i \cdot {\bf S}_{i_B}} |\rangle^{\rm TPQ}_{\beta, N},
\label{eq:local2}
\end{equation}
where the $i$ means the lower site on the vertical $J_D$-bond, $N_m$ is the number of site $i$, which is taken as $N_m=N/4$, and 
$i_T$ and $i_B$ are top and bottom right sites connected to the $i$-th site depicted in the inset of Fig.~\ref{fig:full}. 
For example, we show the results of the temperature dependencies of our two local mirror order parameters in $J/J_D$=0.69 and $N$=32. The result of the empty plaquette in Fig.~\ref{fig:full} is the same one in Fig.~4(b). 
We can see the dominant correlation is the empty one, however, the correlation of the full plaquette pattern is also slightly enhanced at moderate temperatures.
We confirm that the enhancement of the full plaquette pattern could be also understood from the existences of the low-energy excited states exhibiting the full plaquette pattern correlations at the momentum sectors, ($q_x$, $q_y$)=($\pm \frac{\pi}{2}$, 0) and (0, $\pm \frac{\pi}{2}$). We confirm that the X point also has some degenerated excited states exhibiting the full plaquette pattern correlations such as the second and third lower energy excited states.

\bibliography{main.bib}
 
 \end{document}